# An 8 channel parallel transmit system with current sensor feedback for MRI-guided interventional applications

Felipe Godinez[1,2], Raphael Tomi-Tricot[5], Bruno Quesson[8], Matthias Barthel[3], Gunthard Lykowsky[7], Greig Scott[6], Reza Razavi[1,], Joseph Hajnal[1,2] and Shaihan Malik[1,2]

[1] School of Biomedical Engineering and Imaging Sciences, King's College London, London, United kingdom
[2] Centre for the Developing Brain, King's College London, London, United kingdom
[3] Department of Congenital Cardiology, Evelina London Children's Healthcare, Guys and St Thomas' NHS Foundation Trust, London, United Kingdom
[4] Barthel HF-Technik GmbH, Aachen, Germany
[5] MR Research Collaborations, Siemens Healthcare Limited, Frimley, United Kingdom.
[6] Magnetic Resonance Systems Research Laboratory, Department of Electrical Engineering, Stanford University, Stanford, California, USA
[7] RAPID Biomedical GmbH, Rimpar, Germany
[8] Centre de recherche Cardio-Thoracique de Bordeaux/IHU Liryc, INSERM U1045- University of Bordeaux, Pessac, France

E-mail: felipe.godinez@kcl.ac.uk

## Abstract

*Background:* Parallel transmit (pTx) has introduced many benefits to MRI with regard to decreased specific absorption rates and improved transmit field homogeneity, of particular importance in applications at higher magnetic field strengths. PTx has also been proposed as a solution to mitigating dangerous RF induced heating of elongated conductive devices such as those used in cardiac interventions. In this work we present a system that can augment a conventional scanner with pTx, in particular for use in interventional MRI for guidewire safety.

*Methods:* The pTx system was designed to work in-line with a 1.5T MRI while the RF synthesis and imaging control was maintained on the host MR scanner. The add on pTx system relies on the RF transmit signal, unblanking pulse, and a protocol driven trigger from the scanner. The RF transmit was split into multiple fully modulated transmit signals to drive an array of custom transceiver coils. The performance of the 8-channel implementation was tested with regards to active and real-time control of RF induced currents on a standard guidewire, heating mitigation tests, and anatomical imaging in a sheep torso.

*Results:* The pTx system was intended to update RF shims in real-time and it was demonstrated that the safe RF shim could be determined while the guidewire is moved. The anatomical imaging demonstrated that heart anatomy and neighbouring structures can be fully imaged with the pTx system inline.

*Conclusion:* We have presented the design and performance of a real-time feedback control pTx system capable of adding such capabilities to a conventional MRI with the focus guidewire imaging in cardiac interventional MRI applications.

## 1. Introduction

In cardiovascular interventions, magnetic resonance imaging (MRI) provides many benefits over standard x-ray fluoroscopy, from superior soft tissue contrast to advanced functional diagnostics. However, not all interventional devices are MRI safe or can be visualized in MRI. Elongated conductive devices, such as guidewires, that are not ferromagnetic can still cause tissue burns at the proximal tip driven by radiofrequency (RF) induced currents [1], [2], a phenomena caused by the RF coupling between the transmit coil and the interventional device. One way to overcome this effect is to build guidewires from non-metallic materials [3], [4], but this solution sacrifices mechanical performance. A similar approach is to modify the resonance of the device with RF elements [5]. Nevertheless, this requires the adoption of non-standard devices.

To avoid these limitations, some have proposed manipulating the incident electric fields responsible for generating RF induced currents in implanted wires [6]–[8], such that the net electric field tangential to the wire is forced to zero consequently eliminating the potential for localized heating at the tip. This is only possible using parallel transmit (pTx) MR systems which allow spatial manipulation of RF fields by altering the relative amplitude/phase of the individual transmitters. Using a two-channel birdcage coil (now commonly available on 3T MR systems) Eryaman et al [9] demonstrated that a plane through a point on the implanted device with zero electric field could be achieved. Gudino et al [10] demonstrated a similar approach on a guidewire using an eight channel pTx system. Both methods made use of a numerical model to determine the safe RF shim ('RF shim' corresponds to relative amplitude and phase weightings applied to each channel). However, during interventions the RF coupling to the guidewire is constantly changing [11], which requires updating the safe RF shim in real-time to guarantee safety. A proposed solution is to predict the safe RF shim based on a quick measurement instead of a numerical prediction. Etezadi-amoli et al [12] proposed a fast method for computing intrinsically safe RF shims [coined the null modes (NM)] based on measurements made directly on the guidewire using a toroidal current sensor [13]. The same method also produces an unsafe RF shim, coined the coupling mode (CM) that if used carefully can be leveraged to visualize the guidewire [14]. To achieve clinical translation this technique would require a pTx system with a real-time feedback control and be able operate seamlessly with a conventional MRI scanner.

MRI scanners for investigational purposes, with pTx capability, have been built [15], [16]. New commercial 7T scanners can be equipped with a factory pTx system for research purposes. The requirement for a bespoke engineered pTx enabled MRI system for investigation of interventional MRI is a barrier to advancement of this field. An alternative approach is to design a system that can augment a conventional MRI scanner with pTx functionality for research purposes. Custom MRI console systems with such capabilities have been presented [17]–[19]. In both referenced cases the custom console assumes the responsibility of coordinating and delivering RF and gradient timing in coordination with the imaging processes. This approach requires replacing the existing factory MRI console and reimplementing imaging protocols. Alternately, one can augment the RF transmit subsystem only and rely on the host scanner to retain overall control of the imaging experiment. Examples of such 'add-on' pTx systems include a 32-channel system for a 7T scanner [20] and a 64 channel system [21].

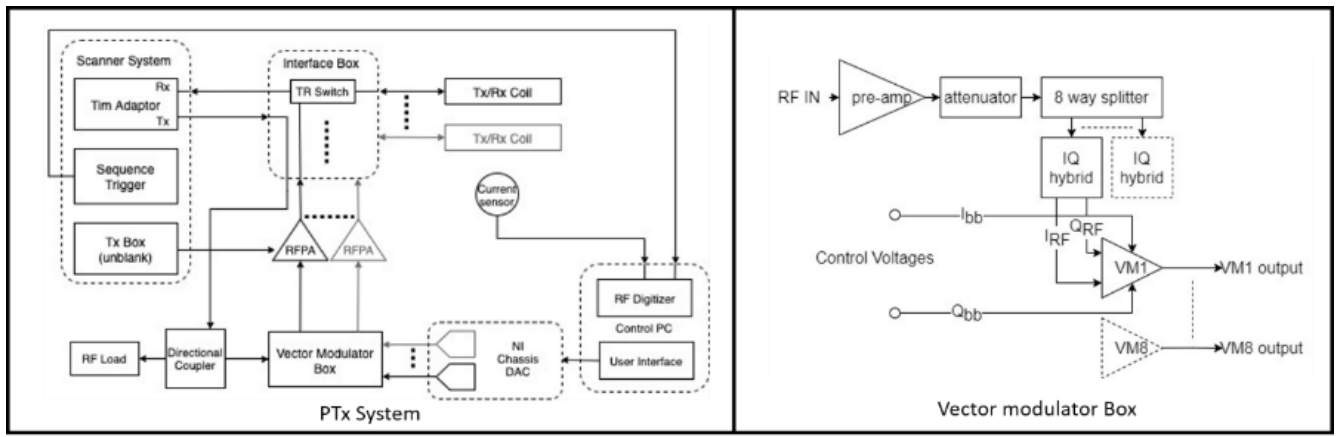

Figure 1 Block schematic of system interface with the MRI scanner and the vector modulator box. The scanner's RF signal was first split 8 ways followed by an IQ power divider before passing through the VMs for amplitude and phase modulation. The auxiliary port on the ADC can be used to record signals from a current sensor, to monitor RF induced currents on a guidewire, or a pickup coil to monitor the coil fields.

This work presents the design and build of a similar 'add-on' pTx system designed to be used for interventional MRI applications, implementing the techniques described in [12], [14]. The key feature of this system is the ability to provide real-time RF current control by applying RF shims that can be updated in real-time during a continuing MR image acquisition. Firstly, the system design is presented in detail, followed by the calibrations used and pertinent methodology, and finally its performance is demonstrated in phantoms and an animal model in the context of interventional MR.

## 2. Methods

### 2.1 System Overview

The proposed system is designed to connect transparently to a conventional MRI scanner through minimally required connections, so that the RF synthesis, sequence control, and data acquisition is kept by the host scanner. Meanwhile the pTx system controls the RF subsystem, by levelling and splitting the RF from the scanner across multiple transmit channels and then independently modulating the amplitude and phase of these (the 'RF shim'), see Figure 1. Periodic measurements from a current sensor are used to update the RF shim during a scan (see section 2.4). The system is intended to perform RF shimming only; it can alter the amplitude/phase of each transmit channel

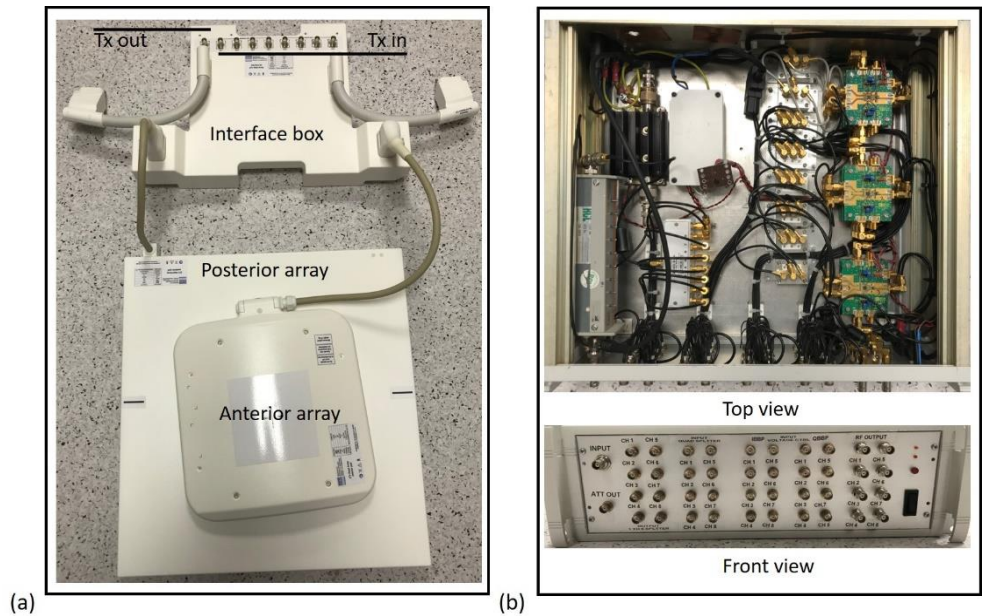

Figure 2 a) The 19" box containing the vector modulators and RF gain components. b) Picture of the coil array used to transmit and receive. There are an anterior 4-coil array and a posterior 4-coil array that mounts into the scanners table. The coils are rectangular loops 50mmx200mm with no overlapping.

for all RF pulses, but not change the 'shape' of the pulse waveforms, which are instead controlled by the scanner. The RF shim is fixed within a repetition time (TR) period.

To achieve these design goals, access to the scanner's synthesized RF source, programmed pulse sequence trigger output, and the RF power amplifier unblank output are required. The pTx system replicates and modulates the RF signal from the scanner, while the pulse sequence trigger initiates the current sensor measurements. The unblank signal is required to toggle the pTx system's RF amplifiers on and off during transmit periods.

A further design requirement is that when 'active' use of the pTx device is not needed, the device should be 'transparent' to the scanner, so that automated power adjustments and pulse sequences from the scanner operate normally. This was accomplished by setting the overall gain of the pTx system to zero relative to the scanner.

*2.2 System Implementation*

The system as implemented has 8 transmit channels and was designed primarily to interface with a Siemens 1.5T MRI system (MAGNETOM Aera, Siemens Healthcare, Erlangen, Germany) using standard connections, though has also been connected to other scanners as detailed below.

*2.2.1   RF coil interface and power amplification*

The system connects to the scanner's local transmit coil socket, known as the Total imaging matrix ('Tim') adaptor, located on the patient bed. The connection is made via an 'interface box' (Figure 1) that connects to the transceiver coil, described in the next section. Use of the local transmit coil socket on the scanner causes the built-in body coil to be detuned and the (high power) RF signal to be routed to the Tim adaptor, and into the interface box. From here the high power signal is sent to a dummy RF load with 3kW capacity, via a directional coupler (C9698-23, 3kW, Werlatone Inc, NY, USA); the attenuated (-40dB) 'forward power' port on this device is the input to our vector modulation stage, described in section 2.2.3.

After modulation the individual channel low-power RF signals are each amplified by an RF power amplifier (RFPA; Barthel HF-Technik GmbH, Aachen, Germany). The RFPAs each have a peak power limit of 1kW and flexible 'energy control' capability allowing for 10% duty cycle at peak power, scaling linearly down to 100W power output at continuous wave operation.

Finally, the amplified and modulated high power RF signals are sent back to the interface box where they are routed via transmit-receive switches to the RF coils. Since the RFPAs and the control computer are located in the control room, all connection are passed through the faraday cage filter panel. The RF load is located in the technical room.

### 2.2.2   Transceiver Coil Array

This work used a custom transceiver surface coil array designed and built by RAPID Biomedical (Rimpar, Germany) who also supplied the interface box (Figure 2a). As well as handling the high-power RF as described above, this interface box routes the received signals via the scanner's standard receiver path. The transceiver has two surface coil arrays (anterior and posterior) each consisting of four 50 mm x 200 mm rectangular loops with no overlap. Nearest and second-nearest-neighbor coil elements were capacitively decoupled.

### 2.2.3   Vector Modulation

The attenuated RF signal coming from the scanner is levelled, using a pre-amp (25dB) (ZHL-6A+, Mini-Circuits, Brooklyn NY, USA) and a variable attenuator (AC701, Pascall Electronics Limited, UK), before being split eight ways (ZCSC-8-1+, Mini-Circuits, Brooklyn NY, USA). Each of the eight signals are passed through a quad hybrid (PSCQ-2-180+, Mini-Circuits, Brooklyn NY, USA) to produce the in-phase (I) and quadrature (Q) signals needed for quadrature modulation using a vector modulator (VM; ADL5390, Analog Devices, Norwood, MA, USA)—evaluation boards were used. Figure 1 shows the block diagram for the network of VMs, which are housed in 3U 19" box, 'VM box' (Figure 2Figure 1b).

A total of 16 independent baseband control voltages [$I_{bb}$, $Q_{bb}$] are required (two per channel) by the VMs, and these are supplied by an 8-bit digital-to-analog converter (DAC) card (PXI-6713, National Instruments, USA) mounted on a PXI chassis (PXI-1033, National Instruments, USA) connected to a PC ('control PC') running Windows 7. The DAC produces a voltage range of ±1.6V; required [$I_{bb}$,$Q_{bb}$] inputs are centered at 0.5V (0V-1V range), such that a setting of [0.5,0.5] produces zero at the VM output.

### 2.2.4   Current sensor

The feedback control loop is achieved by monitoring induced currents via a toroidal coil placed around the segment of the guidewire that is external to the patient's body, close to the entry point. This sensor was built from

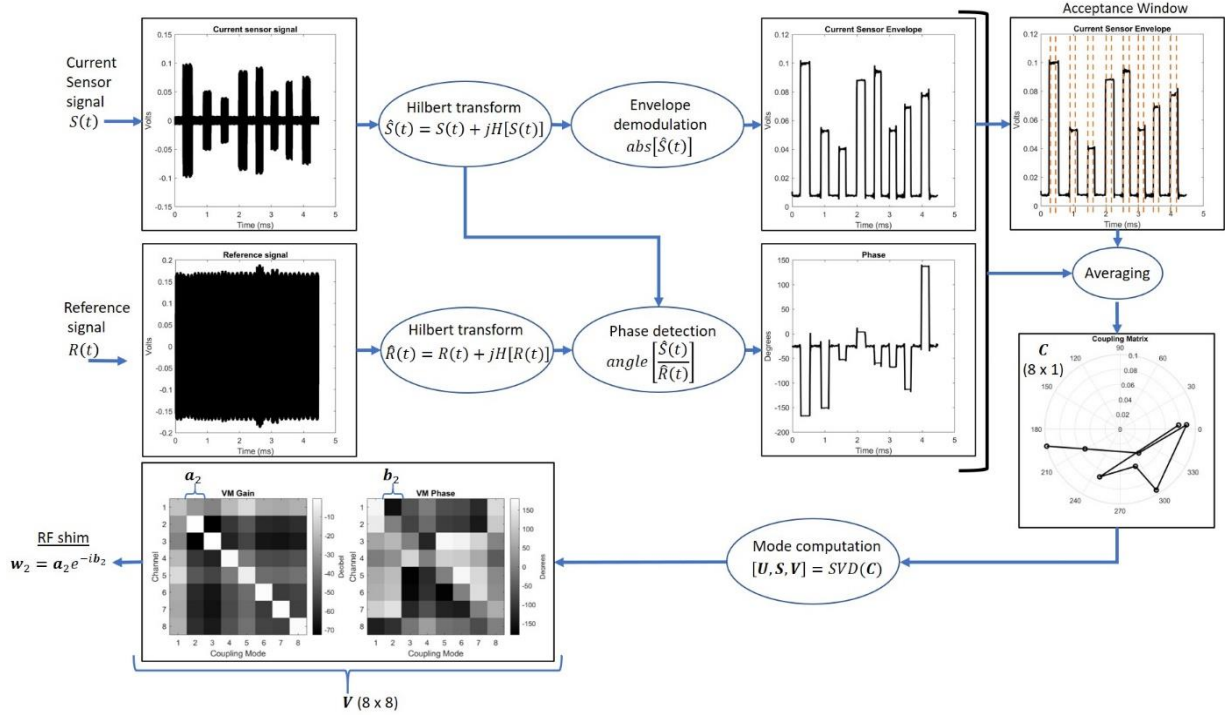

Figure 3 Block diagram of the current sensor *S(t)* and reference *R(t)* signals processing pipeline used to determine the RF shim $w_m$ and coupling modes *V*. The analytical signals $\hat{S}(t)$ and $\hat{R}(t)$ are created by taking the Hilbert transform of the digitized signals *S(t)* and *R(t)*. The elements of the coupling matrix **C** are complex values determined from the envelope demodulation of $\hat{S}(t)$ (VM gain) and the phase detection (VM phase). The coupling modes are computed as the matrix of right-singular vectors of **C** and are numbered such that mode 1 is the high coupling mode while modes 2-8 are null modes.

1 mm diameter transformer wire coiled in the long-axis direction through a 5 mm diameter plastic tube, providing space for the guidewire to pass through. The sensor was wrapped with copper foil to block incident RF energy and a 10m shielded balanced twin-pair line was attached to the coil and connected to a balun. The RF signal from the current sensor is digitized using a four-channel analog-to-digital converter (ADC) (PCI-Gage RMX-161-G40, Dynamic Signals LLC, USA) housed within the control PC. Direct and asynchronous signal demodulation based on the Hilbert transform was used, see section 2.3.1.

*2.3 Control Architecture*

The pTx system performed real-time RF-shim updates by continuously monitoring the current on the guidewire via the current sensor. In a calibration step at the beginning of the running imaging frame, the RF coupling between the guidewire and the individual transmit coils was measured with the current sensor, forming a coupling matrix, ***C***. The calibration was initiated by the pulse sequence trigger followed by a 20ms block RF pulse, during which the VM is used to cycle each channel on/off sequentially, while the current sensor output was recorded (Figure 3). The amplitude and phase of the current measured in each cycle form the elements of ***C*** and were used to compute the coupling modes according to [12]. By applying the singular value decomposition to *C* the coupling modes were determined from the matrix of right-singular vectors *V*. Since this PTx system has 8 channels and 1 current sensor (up to 3 current sensors can be connected), 1 CM and 7 NM were computed. The RF shim was set to either CM or NM by the user and was implemented in the next image frame. The user controls during acquisition are described in section 2.3.2.

*2.3.1    Signal Demodulation*

Figure 3 summarizes the signal processing used to compute the coupling modes from the current sensor signal. A reference signal *R(t)* from the transmit RF was digitized to perform asynchronous and direct demodulation of the current sensor signal *S(t)* [22], [23]. The signal acquisition was under sampled at a sample frequency of 500kHz (RF frequency is 64MHz). Envelope demodulation was performed on *S(t)* by taking the magnitude of the analytical signal created by the Hilbert transform [24] of the digitized real valued signal as follows;

$$\hat{S}(t) = S(t) + jH[S(t)]$$

and

$$m(t) = abs[\hat{S}(t)]$$

Where *m(t)* is the envelope of *S(t)*. The phase $\phi(t)$ of *S(t)* was obtained relative to the reference signal *R(t)* by complex division, as

$$\phi(t) = \tan^{-1}\left[\frac{\hat{S}(t)}{\hat{R}(t)}\right]$$

Since the system is controlled by a PC and not a real-time computer, the switching of the VMs occurs with variable timing. Hence, an edge detection algorithm was employed to set an acceptance window corresponding to each channel section. Once split, the signal was averaged to attain the final value of the amplitude and phase per channel, which makeup the elements of the coupling matrix *C*.

*2.3.2    Software*

The stand-alone system is controlled by a PC running Windows 7. Software to control these devices was written in MatLab R2015b (MathWorks Inc, MA, USA), interfaced via a GUI. Its main functions are:

- i)    Pre-calibration to correct for non-linear response of the VMs
- ii)   Control of VM voltages to each channel for RF-shimming
- iii)  Measurement of the current sensors, including demodulation
- iv)   RF shims computation
- v)    Monitoring and response to the pulse sequence trigger

Since the control software can set the amplitude and phase modulation applied to each channel, it can also control the overall RF power independently of the scanner's regulated transmit level.

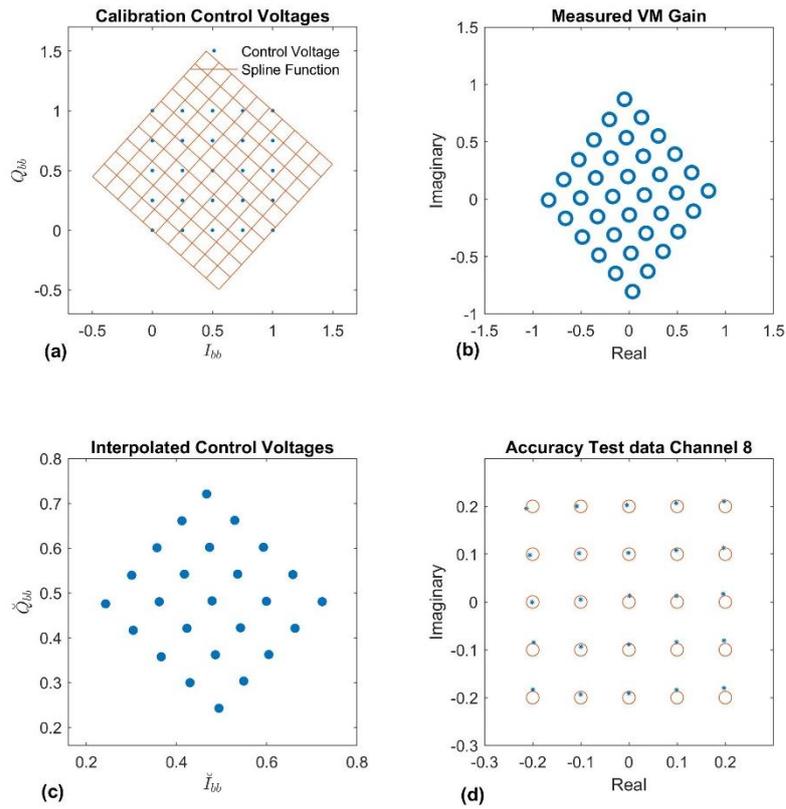

Figure 4 (a) Plot of the thin-plate spline transformation function, mapping the prescribed VM gain measurements, real and imaginary, to the control voltages. The function was determined during calibration using a 5x5 constellation of points. (b) The measured VM gain at the calibration control voltages shown in (a). (c) The interpolated control voltages determined after the calibration for prescribed target VM gains. (d) The target VM gain (open circles) points plotted with the measured VM gain points (solid dots).

*2.4 Vector Modulator Calibration*

VMs can produce a non-linear response because of analog mixer imperfections, and isolation errors in the splitting and combining networks at the inputs or outputs. Fortunately, this can be corrected via a feed-forward correction approach [25], which requires prior knowledge of the non-linear distortions in the output gain and phase. Using this approach, the VMs were calibrated at the start of each session. The calibration consisted in determining a pre-distortion function; in this work a thin-plate spline function [26], [27] was used, which estimates the pre-distorted control voltages $[\check{I}_{bb}, \check{Q}_{bb}]$ that produce a desired VM output. The function for each VM was determined experimentally by measuring the VM gain and phase for an input Cartesian grid of $[I_{bb}, Q_{bb}]$ values ranging from 0V to 1V.

The fully automated calibration was done using a benchtop signal generator as a 64MHz RF source and the ADC card to measure the amplitude/phase of the modulated signal. Subsequently, the thin-plate spline function was fit to the measured data. The thin-plate function used is,

$$f(\mathbf{x}) = \sum_{j=1}^{n-3} \left( |\mathbf{x} - \mathbf{c}_j|^2 \log|\mathbf{x} - \mathbf{c}_j|^2 \right) a_j + x_1 a_{n-2} + x_2 a_{n-1} + a_n$$

Where $f(\mathbf{x})$ is a function that interpolates the control voltages $[\check{I}_{bb}, \check{Q}_{bb}]$ given the real and imaginary parts of the desired (the RF shim) VM gain and phase, handled as a complex point $\mathbf{x}$ ($x_1$=*real*, $x_2$=*imaginary*). The variable *n* is the number of data points used during the calibration measurement, and $\mathbf{c}_j$ is the complex Cartesian coordinate (real, imaginary) of the measured VM gain/phase. The polynomial coefficients $a_n, a_{n-1}, a_{n-2}, a_j$ are computed using the **tpaps(c,y)** function in MatLab which finds the minimizer *f* using the following cost function

$$E(f) = \sum_j |\mathbf{y}(j) - f(\mathbf{c}(:,j))|^2.$$

where **y** is a vector of test control voltages [$I_{bb}$,$Q_{bb}$] chosen for the calibration to measure the VM gain, **c** (Figure 4a-b). Using the function, *f*, the [$\check{I}_{bb}$,$\check{Q}_{bb}$] needed for a prescribed RF shim can be determined by interpolating between the measured data points (Figure 4c). The function was determined with a 5x5 Cartesian grid of control voltages ranging from 0V-1V, the VM input range.

The accuracy of the calibration was tested by prescribing a set of RF shims and then measuring the resulting output for each channel (Figure 4c-d). The average absolute RMS error between the prescribed RF shim and measured gain was computed for each channel using all the points in the Cartesian grid.

*2.5 Real-time control*

The system's ability to update the coupling modes in real-time was tested using a 3T Achieva MRI scanner (Philips, The Netherlands) with a built-in 8-channel pTx transceiver array body coil [15]. This system was used for this test because the large built-in pTx body coil allowed a guidewire to be freely moved within it to test for control of induced currents over a range of positions. In this experiment the add-on pTx system was connected in a similar manner as described above for a Siemens 1.5T system, but directly to the pTx body coil present on the 3T system, rather than the local array coil which was not used. The modulated transmit signals were amplified using the 3T system's RFPAs. To operate at 3T the hybrid splitters were swapped for counterparts at the desired frequency.

A straight guidewire (90cm long nitinol guidewire, Terumo Corporation, Japan) was placed axially in the scanner's bore in air, instrumented with a toroidal current sensor. The test used a turbo spin echo (TSE) sequence running at 100% reported SAR with TR=4s, modified by adding the coupling matrix measurement block (30ms) prior to each echo train. The software of the add-on pTx system could be set to either use this information to update the coupling mode calculation once per image (4s) or ignore the new measurements and fix at a previous value.

A starting coupling matrix measurement was made with the guidewire placed at the left-hand edge of the scanner bed, position-A. The guidewire was then moved to the right-hand bed edge, position-B, without updating the modes (i.e. real-time control disabled). After some acquisitions at position-B the real-time control was enabled to dynamically control the measured guidewire currents. The guidewire was then moved between position-A and position-B while dynamically updating the coupling measurements.

*2.6 MR imaging*

Imaging performance was evaluated on the Siemens 1.5T MRI system mainly focused on torso imaging emulating interventional MR both in phantoms and ex vivo.

*2.6.1 Heating tests*

To demonstrate the efficacy of the pTx system to mitigate guidewire tip heating, a standard 0.89mm diameter guidewire (RF+GA35153M, Terumo Corporation, Japan) cut to 90cm length and stripped 3mm at the tip, was equipped with a fiber-optic temperature probe (LumaSense Technologies, Inc. USA). The guidewire was placed in a poly-acrylic acid gel phantom built according to [28] with the dimensions 45cm x 36 cm x 9cm. A high SAR steady-state free precession sequence was used to induce heating [14].

*2.6.2 Anatomical imaging*

Anatomical images were obtained on a sheep torso to qualitatively evaluate system performance under realistic experimental conditions. Images were acquired with a 3D GRE sequence with an isotropic spatial resolution of 0.8mm and with a TR/TE 389.77ms/1.82ms. The pTx system was set with an RF shim of equal amplitude per channel and phased to achieve a nominal circular polarized mode; no guidewire was used. All procedures involving animals were approved by the Animal Research Ethics Committee (CEEA 50 - France) and performed in accordance with the European rules for animal experimentation.

**3. Results**

*3.1 Vector modulator calibration*

The observed distortion in the VM gain can be characterized as a rotation with a translation Figure 4. After calibration the average RMSE across VM channels was 11% (min1.1% and max 28.0%) of the dynamic range tested.

*3.2 Real-Time Control*

Figure 5 shows the real-time performance of the PTx system while moving a guidewire in air while switching between CM and NM which should register high and low coupling, respectively. When the real-time control is disabled the measured coupling is similar for both the CM (0.49V) and NM (0.39V). When it is enabled the measurements follow the expected response from the CM (1.05V) and NM (0.05V), even when the guidewire is in motion.

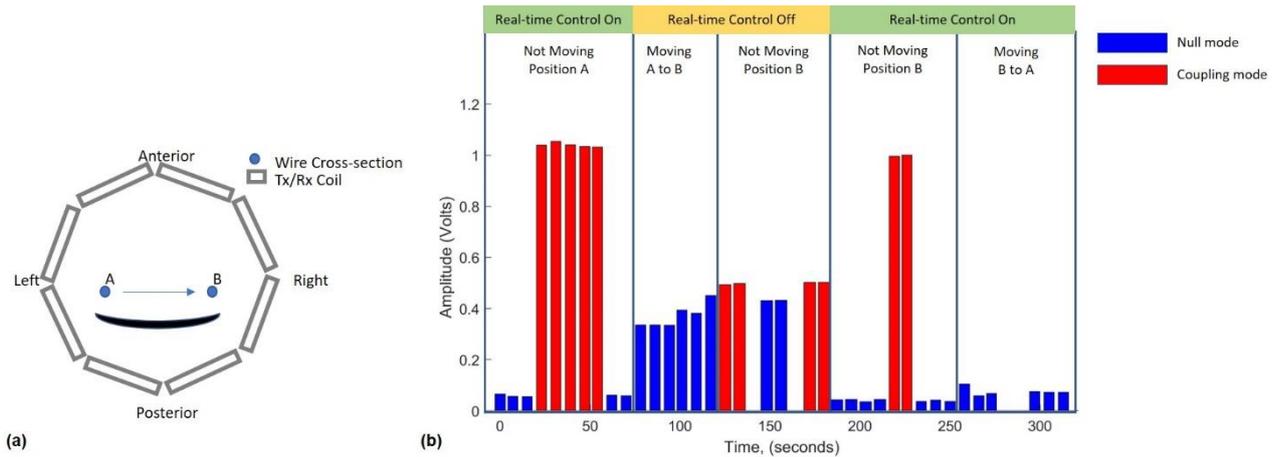

Figure 5 Plot of current measurements on the guidewire while waving in air during real-time control. a) shematic of guidewire positions in pTx body coil. b) results of real-time test. The guidewire starts at position A (left-hand side of scanner bore) and without moving the guidewire coupling modes are determined and engaged. After 70s the wire is moved from position A to B (right-hand side of scanner bore) with the real-time control off, but with the null mode still engaged, determined at position A. At 120s the null mode and coupling mode are switched while the real-time controller is off. Then at 187s the real-time controller is turned on and new coupling modes are measured. Finally the 258s the guidewire is moved back to position A while the real-time control is enabled while the null mode is engaged. In this configuration the real-time accurately updates the null mode as the RF coupling to the guidewire is actively changing with the motion.

*3.3 Heating test*

Figure 6 shows the measured temperature at the guidewire tip in a phantom experiment while scanning with either the CM or NM at equivalent RF power. The measured temperature increase was reduced from 23.89°C in CM to 1.65°C in NM, demonstrating strong suppression of induced currents.

*3.4 MR PTx Imaging*

The sheep torso image shown in Figure 7 demonstrates that standard anatomical imaging is possible while using the pTx system in CP mode. During this scan the scanner completed its standard power scaling procedure and reported a reference voltage of 217v (maximum allowed was 234v). This demonstrates that when not actively controlled the system can operate under the full control of the host MR scanner.

**4. Discussion**

An auxiliary pTx system, intended for interventional MRI, was assembled, and proven to add parallel transmit capabilities to a non-pTx scanner. Its key function was to perform RF shimming, operating independently from the host scanner during imaging with real-time feedback control based on concurrent current measurements. The

arXiv:2103.10399

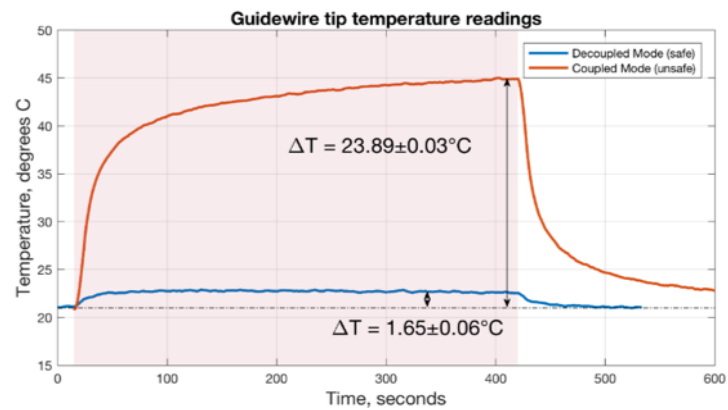

Figure 6 Temperature measurements at the tip of the guidewire during the heating SSFP sequence while using either the null or coupling mode. Heating is shown the increase during the coupling mode and minimized with the null mode.

results indicate that when set to standard CP mode the device can be used to produce images using standard protocols. Active control of induced currents on inserted guidewires was also demonstrated, with strong attenuation of temperature increases and active control.

This add-on pTx system was primarily designed for a 1.5T application but was also connected to a 3T scanner with non-standard body coil. In that case some narrow-band RF components (hybrid splitters and RFPAs) had to be swapped to the appropriate frequency. If such a system is required to work across frequencies, then broadband components could be selected where available – this was not a design requirement for the device as presented.

A limitation for this demonstration device is that specific absorption rate (SAR) monitoring was not implemented, which would be a requirement for progression of this work to human imaging. The amplifiers used in this work did have integrated directional couplers and a digitization system that reported average forward and reflected power based on a short time window, and this could form the basis of a future SAR monitoring framework.

The absence of a real-time operating system imposes a non-deterministic performance on the system, barring the direct shaping of RF pulses. Although dynamic shimming is possible, at base band signals with a bandwidth less than 1.85kHz, it would not be sufficient to handle complete RF pulse envelope modulation [20], which could demand bandwidths of 100kHz or more. Since the DAC card in this build has a sampling rate of up to 769 kS/s per channel, the bandwidth bottle neck is the operating system priority control. The average interval between VM control voltage updates was 274±50μs.

The presented pTx system is intended for further development of interventional MRI using endovascular guidewires. In this design a single toroidal current sensor was used to measure currents. In future this could be expanded to multiple sensors or other sensors that provide relevant real-time information.

arXiv:2103.10399

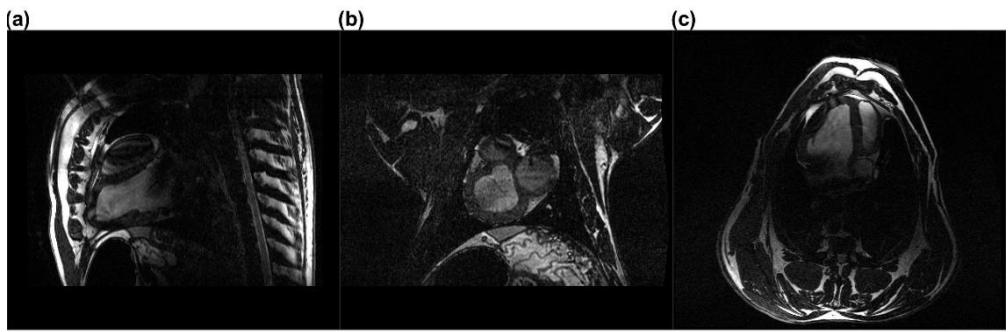

Figure 7 Ex vivo MR images of the sheep torso. (a) Sagittal, (b) coronal, and (c) axial plane of torso MR image acquired with a 3D GRE sequence.

The demonstrated anatomical imaging revealed lower signal at the centre of the field of view because of the use of surface coils that have strongly surface-weighted sensitivity. This issue could be addressed by improved RF coil design; this is outside the scope of this work, which focused on the pTx control system development. It should also be noted that for interventional work the drop in standard imaging performance may still be acceptable should the system allow clinicians to perform hitherto unavailable procedures. This will be the target of future work.

5. Conclusion

We have built an add-on pTx system that can be used to augment conventional MRI scanners to add pTx functionality, for interventional MRI. The system is capable of RF shimming and functions transparently to the host MR system.

**References**

[1] M. F. Dempsey, B. Condon, and D. M. Hadley, "Investigation of the factors responsible for burns during MRI," *J. Magn. Reson. Imaging*, vol. 13, no. 4, pp. 627–631, 2001, doi: 10.1002/jmri.1088.

[2] A. Buecker, "Safety of MRI-guided vascular interventions," *Minim. Invasive Ther. Allied Technol.*, vol. 15, no. 2, pp. 65–70, 2006, doi: 10.1080/13645700600640717.

[3] R. Mekle, E. Hofmann, K. Scheffler, and D. Bilecen, "A polymer-based MR-compatible guidewire: A study to explore new prospects for interventional peripheral magnetic resonance angiography (ipMRA)," *J. Magn. Reson. Imaging*, vol. 23, no. 2, pp. 145–155, 2006, doi: 10.1002/jmri.20486.

[4] A. Massmann, A. Buecker, and G. K. Schneider, "Glass-fiber-based MR-safe guidewire for MR imaging-guided endovascular interventions: In vitro and preclinical in vivo feasibility study," *Radiology*, 2017.

[5] M. E. Ladd and H. H. Quick, "Reduction of resonant RF heating in intravascular catheters using coaxial chokes," *Magn. Reson. Med.*, vol. 43, no. 4, pp. 615–619, 2000, doi: 10.1002/(SICI)1522-

2594(200004)43:4<615::AID-MRM19>3.0.CO;2-B.

[6]   C. E. McElcheran, B. Yang, K. J. T. Anderson, L. Golenstani-Rad, and S. J. Graham, "Investigation of parallel radiofrequency transmission for the reduction of heating in long conductive leads in 3 Tesla magnetic resonance imaging," *PLoS One*, vol. 10, no. 8, pp. 1–21, 2015, doi: 10.1371/journal.pone.0134379.

[7]   Y. Eryaman *et al.*, "A simple geometric analysis method for measuring and mitigating RF induced currents on Deep Brain Stimulation leads by multichannel transmission/reception," *Neuroimage*, vol. 184, no. May 2018, pp. 658–668, 2019, doi: 10.1016/j.neuroimage.2018.09.072.

[8]   L. Winter *et al.*, "Parallel transmission medical implant safety testbed: Real-time mitigation of RF induced tip heating using time-domain E-field sensors," *Magn. Reson. Med.*, vol. 84, no. 6, pp. 3468–3484, 2020, doi: 10.1002/mrm.28379.

[9]   Y. Eryaman, B. Akin, and E. Atalar, "Reduction of Implant RF Heating Through Modification of Transmit Coil Electric Field," vol. 1313, pp. 1305–1313, 2011, doi: 10.1002/mrm.22724.

[10]  N. Gudino *et al.*, "Parallel transmit excitation at 1.5 T based on the minimization of a driving function for device heating," *Med. Phys.*, vol. 42, no. 1, pp. 359–371, 2015, doi: 10.1118/1.4903894.

[11]  C. Armenean, E. Perrin, M. Armenean, O. Beuf, F. Pilleul, and H. Saint-Jalmes, "RF-induced temperature elevation along metallic wires in clinical magnetic resonance imaging: Influence of diameter and length," *Magn. Reson. Med.*, vol. 52, no. 5, pp. 1200–1206, 2004, doi: 10.1002/mrm.20246.

[12]  M. Etezadi-Amoli, P. Stang, A. Kerr, J. Pauly, and G. Scott, "Controlling radiofrequency-induced currents in guidewires using parallel transmit," *Magn. Reson. Med.*, vol. 74, no. 6, pp. 1790–1802, Dec. 2015, doi: 10.1002/mrm.25543.

[13]  M. G. Zanchi, R. Venook, J. M. Pauly, and G. C. Scott, "An optically coupled system for quantitative monitoring of MRI-induced RF currents into long conductors," *IEEE Trans. Med. Imaging*, vol. 29, no. 1, pp. 169–178, Jan. 2010, doi: 10.1109/TMI.2009.2031558.

[14]  F. Godinez, G. Scott, F. Padormo, J. V. Hajnal, and S. J. Malik, "Safe guidewire visualization using the modes of a PTx transmit array MR system," *Magn. Reson. Med.*, vol. 83, no. 6, pp. 2343–2355, 2020, doi: 10.1002/mrm.28069.

[15]  P. Vernickel *et al.*, "Eight-channel transmit/receive body MRI coil at 3T," *Magn. Reson. Med.*, vol. 58, no. 2, pp. 381–389, Aug. 2007, doi: 10.1002/mrm.21294.

[16]  F. X. Hebrank *et al.*, "Parallel Transmission ( pTX ) Technology * MR Imaging with an 8-Channel RF Transmit Array," *Magnetom flash*, pp. 75–79, 2007.

[17]  P. Stang, S. Conolly, J. Pauly, and G. Scott, "MEDUSA: A Scalable MR Console for Parallel Imaging," 2007.

[18]  P. P. Stang, S. M. Conolly, J. M. Santos, J. M. Pauly, and G. C. Scott, "Medusa: A scalable MR console using USB," *IEEE Trans. Med. Imaging*, vol. 31, no. 2, pp. 370–379, 2012, doi: 10.1109/TMI.2011.2169681.

[19]  A. Ang, S. Obruchkov, and R. Dykstra, "Construction of an open PXIe based scalable MRI console," in *Proc. Intl. Soc. Mag. Reson. Med.*, 2018, vol. 26, pp. 18–20.

[20]  S. Orzada *et al.*, "A 32-channel parallel transmit system add-on for 7T MRI," *PLoS One*, vol. 14, no. 9, p. e0222452, Sep. 2019, doi: 10.1371/journal.pone.0222452.

arXiv:2103.10399

[21]   K. Feng, N. A. Hollingsworth, M. P. McDougall, and S. M. Wright, "A 64-channel transmitter for investigating parallel transmit MRI," *IEEE Trans. Biomed. Eng.*, vol. 59, no. 8, pp. 2152–2160, 2012, doi: 10.1109/TBME.2012.2196797.

[22]   P. Pérez, A. Santos, and J. J. Vaquero, "Potential use of the undersampling technique in the acquisition of nuclear magnetic resonance signals," *Magn. Reson. Mater. Physics, Biol. Med.*, vol. 13, no. 2, pp. 109–117, 2001, doi: 10.1016/S1352-8661(01)00137-5.

[23]   D. Gareis, T. Neuberger, V. C. Behr, P. M. Jakob, C. Faber, and M. A. Griswold, "Application of Undersampling Technique for the Design of an NMR Signals Digital Receiver," *Concepts Magn. Reson. Part B*, vol. 29B, no. 1, pp. 20–27, 2006, doi: 10.1002/cmr.b.

[24]   L. Xiaozhou, W. Qianjuan, L. Jingjing, and J. Haibin, "Signal Demodulation Algorithm based on Hilbert Transform and its applicaTion in Aircraft Power Supply System Characteristic Parameters Testing," *2019 14th IEEE Int. Conf. Electron. Meas. Instruments, ICEMI 2019*, pp. 1549–1554, 2019, doi: 10.1109/ICEMI46757.2019.9101588.

[25]   P. Tošovský and D. Valúch, "Improvement of RF vector modulator performance by feed-forward based calibration," *Radioengineering*, vol. 19, no. 4, pp. 627–632, 2010.

[26]   F. L. Bookstein, "Principal Warps: Thin-Plates Splines and the decompostion of deformations," *IEEE Trans. Pattern Anal. Mach. Intell.*, vol. 11, no. 6, pp. 567–585, 1989, doi: 10.1109/34.24792.

[27]   F. Padormo, S. J. Malik, G. Mens, and J. V. Hajnal, "A method for calibrating multi-channel RF systems," in *Proc. Intl. Soc. Mag. Reson. Med.*, 2011, vol. 19, p. 3858.

[28]   ASTM standard F 2182-2002a, "Standard test method for measurement of radio frequency induced heating near passive implants during magnetic resonance imaging," *ASTM Int.*, no. August, pp. 1–14, 2011, doi: 10.1520/F2182-11A.1.7.